\documentclass[runningheads,a4paper]{llncs}
\usepackage[utf8]{inputenc}
\usepackage[T5]{fontenc}
\usepackage{graphicx,booktabs,array,tabularx,longtable,amsmath,amssymb,tikz,float,needspace,etoolbox,placeins}
\usepackage[table]{xcolor}
\usetikzlibrary{arrows.meta,positioning}
\usepackage{hyperref}
\usepackage{orcidlink}
\usepackage{marvosym,accsupp}
\DeclareRobustCommand{\correspondingmark}{%
  \BeginAccSupp{method=hex,unicode,ActualText=002827090029}%
  {\large(\Letter)}\EndAccSupp{}%
}
\DeclareMathSizes{7.4}{7.4}{5}{5}
\makeatletter
\renewcommand{\LT@makecaption}[3]{%
  \LT@mcol\LT@cols c{\hbox to\z@{\hss
    \parbox[t]\LTcapwidth{\small
      \sbox\@tempboxa{#1{\textbf{#2.}\ }#3}%
      \ifdim\wd\@tempboxa>\hsize
        #1{\textbf{#2.}\ }#3%
      \else
        \hbox to\hsize{\hfil\box\@tempboxa\hfil}%
      \fi\endgraf\vskip\baselineskip}%
    \hss}}}
\makeatother
\hypersetup{colorlinks=true,linkcolor=blue,citecolor=blue,urlcolor=blue}
\definecolor{sigred}{HTML}{B3242D}
\definecolor{softblue}{HTML}{EFF5F7}
\definecolor{softgreen}{HTML}{EFF5F0}
\definecolor{tableband}{HTML}{EFF3F4}
\definecolor{tablelabel}{HTML}{E8EDF7}
\definecolor{winblue}{HTML}{315F82}
\definecolor{tiegray}{HTML}{CBD4D9}
\definecolor{lossred}{HTML}{B5474E}
\newcommand{\sigfive}{\makebox[0pt][l]{\kern2pt\textcolor{sigred}{\textsuperscript{$\dagger$}}}}
\newcommand{\sigone}{\makebox[0pt][l]{\kern2pt\textcolor{sigred}{\textsuperscript{$\ddagger$}}}}
\newsavebox{\normalmetric}
\newsavebox{\boldmetric}
\newcommand{\best}[1]{\mathord{\hbox{\begingroup\sbox{\normalmetric}{$#1$}\sbox{\boldmetric}{$\boldsymbol{#1}$}\makebox[\wd\normalmetric][c]{\resizebox{\wd\normalmetric}{!}{\usebox\boldmetric}}\endgroup}}}
\begin{document}
\title{Structure and Lineage Evidence for Vietnamese Statutory Article Retrieval on DRiLL}
\titlerunning{Structure and Lineage Evidence for Vietnamese Legal Retrieval on DRiLL}
\author{Duc-Vu Nguyen\inst{1,2}\orcidlink{0000-0003-0072-2524}\,\correspondingmark\and Ngan Luu-Thuy Nguyen\inst{1,2}\orcidlink{0000-0003-3931-849X}}
\authorrunning{Duc-Vu Nguyen et al.}
\institute{University of Information Technology, Ho Chi Minh City, Vietnam\\
\and Vietnam National University, Ho Chi Minh City, Vietnam\\
\email{\{vund,ngannlt\}@uit.edu.vn}}
\maketitle

\begin{abstract}
We study Vietnamese statutory Article retrieval on DRiLL using features for provision identity and local form, same-law support, and citation and lineage evidence. Across five random seeds, selectors sharing candidates, listwise Qwen evidence, and learner settings improve official $F_2$ from 71.77\% to 72.91\%, with the largest descriptive gain on two-Article questions. Added features account for 18.20\% of the proposed selector's total decision-tree split gain, measured as training-objective reduction, mainly from same-law and citation evidence. Cases reveal base provisions and amendments, companion recovery, distractor removal, and failures. These findings motivate research on other datasets, representations, and learners without establishing individual feature effects or requiring LightGBM, hand-crafted scores, or explicit graphs.
\end{abstract}

\section{Introduction}
Statutory Article retrieval connects questions with the provisions needed to answer them~\cite{louis2022bsard}. Similar wording may miss complementary rules or amendments. Vietnamese statutes expose structure through \emph{Điều} (Article), \emph{Khoản} (Clause), \emph{Điểm} (Point), instrument identifiers, and references. DRiLL includes questions requiring multiple Articles~\cite{vuong2025drill}, making it suitable for examining relationships beyond individual similarity scores.

Our research question is: \emph{Does DRiLL contain structural and lineage-related evidence that can help retrieval beyond conventional similarity?} Structure includes legal-unit form, instrument membership, and reference neighborhoods. Lineage-related cues connect base provisions with amendments, guidance, or introduced units. We examine these cues through a retrieval method.

We follow EDM's DRiLL retrieval setting, conventional features, and staged design~\cite{pham2025edm}, simplifying the LLM-dependent stages to reduce LLM calls. We then add legal evidence to the internal controls. Final selectors share candidates, listwise judgments, and learner settings. Five-run results, split-gain importance, and cases characterize DRiLL evidence through LightGBM and explicit features. Future work can explore other learners, representations, and datasets.

\section{Related Work}
\noindent\textbf{Vietnamese retrieval systems.}
DRiLL systems combine lexical and semantic matching with richer Article context. Hieu and Do use hybrid retrieval and a cross-encoder, with reconstructed titles and hard-negative training~\cite{hieu2025twostage}. Their design enriches individual candidates with document context. Hoai et al. use LLM augmentation and hierarchical chunking to form coherent retrieval units~\cite{hoai2025chunking}. This addresses the internal organization of long legal texts. ViDRILL combines sparse and dense retrieval with semantic reranking, dual-level chunking, and hard-negative mining~\cite{tran2025vidrill}. These pipelines improve matching and candidate quality. We examine additional evidence connecting candidates with other provisions.

\noindent\textbf{LLM evidence and staged ranking.}
HyDE embeds a generated hypothetical document to retrieve real corpus passages~\cite{gao2023hyde}. This enriches the query view, while retrieved legal texts remain the evidence for answering. EDM combines hybrid retrieval, learned ranking, and LLM relevance judgments~\cite{pham2025edm}. It also uses citation expansion and combines matching scores with document-form cues. Our internal controls reuse this feature design while simplifying the LLM stages. The comparison examines added legal evidence under shared final-selection inputs. We distinguish logical LLM evaluations implied by these designs from measured latency or monetary cost.

\noindent\textbf{Structural and lineage context.}
G-DSR incorporates legislative structure into dense retrieval through a graph neural network~\cite{louis2023finding}. This captures dependencies beyond an isolated Article. LegisSearch combines a legislative graph, contextual embeddings, and LLM query expansion for Italian legislation~\cite{colombo2025legissearch}. Its typed citation and amendment relations provide a precedent for representing lineage. These approaches motivate examining the direction and type of legal connections. Extracted DRiLL links remain potential relevance cues. A relation may identify a missing companion or a topical distractor, so its value depends on the query and other candidates available for selection. We study same-instrument support, reference neighborhoods, and amendment-related patterns through explicit features and cases.

\section{Task and DRiLL Data}
For question $q$, let $\mathcal A$ be the Article universe and $G_q\subseteq\mathcal A$ the set of distinct gold Article IDs. A system produces ranking $\pi_q$ and selected set $S_q$. Per-question precision and recall are $P_q=|S_q\cap G_q|/|S_q|$ and $R_q=|S_q\cap G_q|/|G_q|$, with zero precision for an empty selection. DRiLL computes its official aggregate after averaging precision and recall across questions:
\begin{equation}
 \bar P=\frac1Q\sum_qP_q,\qquad \bar R=\frac1Q\sum_qR_q,\qquad F_2=\frac{5\bar P\bar R}{4\bar P+\bar R}.
 \label{eq:f2}
\end{equation}
First-ranker R@$k$ is mean per-question gold recall within its first $k$ Articles, reported at $k=10$ and $100$. Descriptive slices use $F_{2,q}=5P_qR_q/(4P_q+R_q)$, with zero when both inputs are zero. Its mean differs from Eq.~\eqref{eq:f2}.

We use the published DRiLL corpus and partitions~\cite{vuong2025drill}: 59,636 Articles and 2,190 training, 312 development, and 627 test questions (Table~\ref{tab:data}). Development and test correspond to the organizer's public-test and private-test releases. All feature configurations and hyperparameters are tuned on development data. After deduplicating gold IDs, test contains 474 single-Article, 121 two-Article, and 32 larger sets. The 153 multi-Article questions constitute 24.4\%.

\begin{table}[htb]\centering\small
\caption{Question counts in DRiLL. Multi-Article counts are derived using at least two distinct gold Article IDs per question.}
\label{tab:data}\setlength{\tabcolsep}{6pt}
\begin{tabular}{l>{\columncolor{tableband}}r r>{\columncolor{tableband}}r}\toprule
\textbf{Study split} & \textbf{Questions} & \textbf{Multi-Article} & \textbf{Published partition}\\\midrule
Train & 2,190 & 544 & Training\\\hline
Development & 312 & 79 & Public test\\\hline
Test & 627 & 153 & Private test\\\bottomrule
\end{tabular}
\end{table}

\section{Retrieval with Legal-Inspired Features}
\subsection{Pipeline and the scope of the controls}
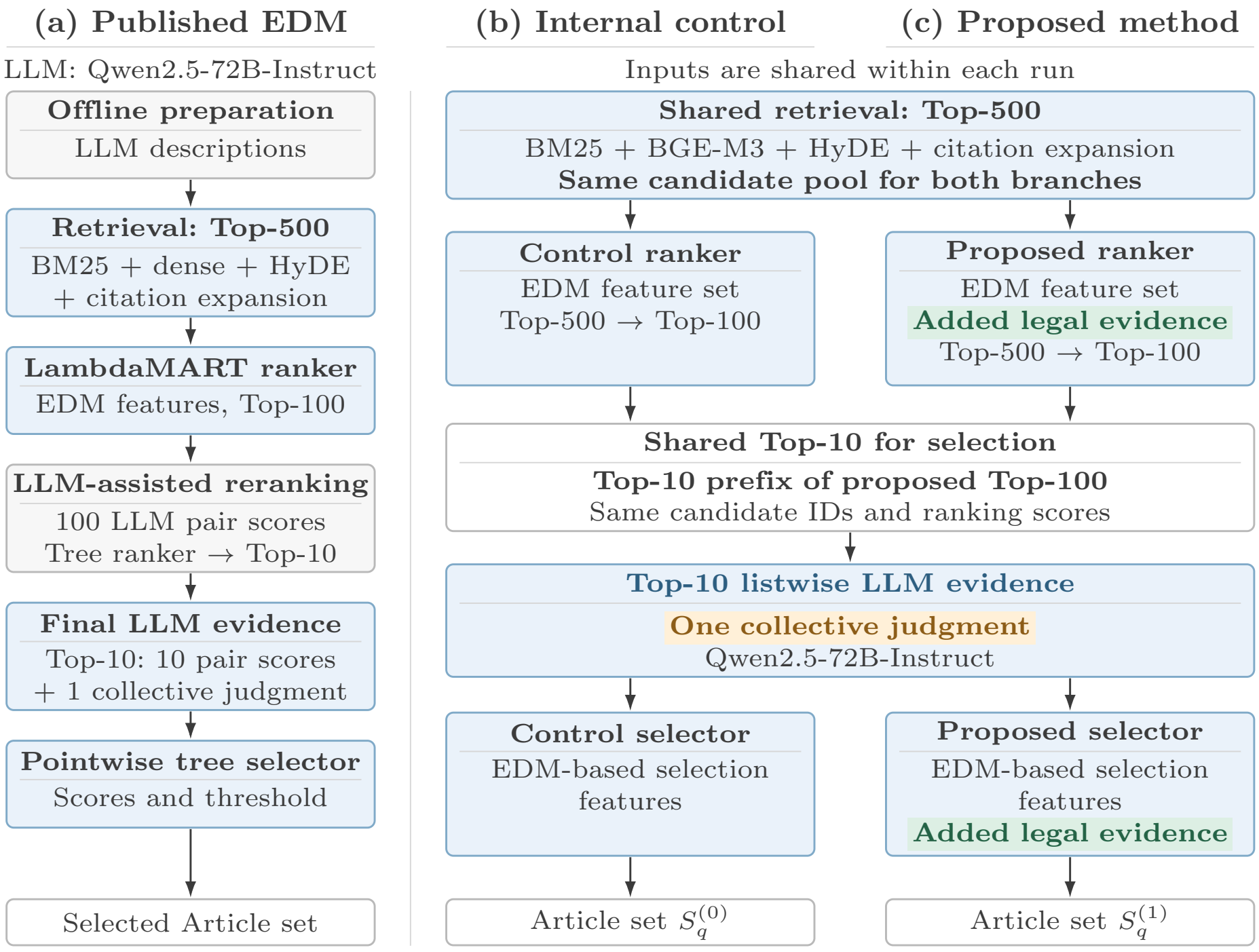
\begin{figure}[!t]
\centering
\resizebox{\linewidth}{!}{
\begingroup%
\definecolor{drillInk}{HTML}{3B3B3B}%
\definecolor{drillLine}{HTML}{B9B9B9}%
\definecolor{drillRule}{HTML}{D6D6D6}%
\definecolor{drillGray}{HTML}{F7F7F7}%
\definecolor{drillLegalFill}{HTML}{DDEFE4}%
\definecolor{drillLegalInk}{HTML}{246346}%
\definecolor{drillBlueLine}{HTML}{80AAC8}%
\definecolor{drillBlueFill}{HTML}{EAF2FA}%
\definecolor{drillBlueInk}{HTML}{2D5E80}%
\definecolor{drillAdaptFill}{HTML}{FFF0D8}%
\definecolor{drillAdaptInk}{HTML}{8C5C17}%
\newcommand{\drillBadge}[3]{\begingroup\setlength{\fboxsep}{1.3pt}%
  \colorbox{#1}{\textcolor{#2}{\textbf{#3}}}\endgroup}%
\newcommand{\drillModule}[8]{%
  \node[drillBox,#6,minimum width=#4cm,minimum height=#5cm]
    (#1) at (#2,#3) {};
  \node[drillTitle] at ([yshift=-.20cm]#1.north) {\mbox{#7}};
  \draw[drillDivider]
    ([xshift=.13cm,yshift=-.38cm]#1.north west)--
    ([xshift=-.13cm,yshift=-.38cm]#1.north east);
  \node[drillBody,anchor=north,text width=\dimexpr #4cm-8pt\relax]
    at ([yshift=-.46cm]#1.north) {#8};
}%
\begin{tikzpicture}[
  x=1cm,y=1cm,
  font=\normalfont\rmfamily\fontsize{7.4}{8.8}\selectfont,
  text=drillInk,
  every node/.style={inner sep=0pt,outer sep=0pt,align=center},
  drillBox/.style={draw=drillLine,fill=drillGray,line width=.55pt,
    rounded corners=2pt},
  drillBlue/.style={draw=drillBlueLine,fill=drillBlueFill},
  drillTitle/.style={font=\normalfont\rmfamily\bfseries\fontsize{7.4}{8.8}\selectfont},
  drillBody/.style={font=\normalfont\rmfamily\fontsize{7.4}{8.8}\selectfont},
  drillPanel/.style={font=\normalfont\rmfamily\fontsize{8.6}{10}\selectfont\bfseries},
  drillDivider/.style={draw=drillRule,line width=.35pt},
  drillFlow/.style={-{Latex[length=1.6mm,width=1.1mm]},draw=drillInk,
    line width=.55pt,shorten <=.4pt,shorten >=.4pt}
]

\node[drillPanel] at (1.80,.03) {(a) Published EDM};
\node[drillPanel] at (6.10,.03) {(b) Internal control};
\node[drillPanel] at (10.40,.03) {(c) Proposed method};
\draw[drillDivider] (0,-.19)--(3.60,-.19);
\draw[drillDivider] (4.30,-.19)--(7.90,-.19);
\draw[drillDivider] (8.60,-.19)--(12.20,-.19);
\node[drillBody] at (1.80,-.43) {LLM: Qwen2.5-72B-Instruct};
\node[drillBody] at (8.25,-.43) {Inputs are shared within each run};
\draw[drillDivider] (3.95,-.62)--(3.95,-8.98);

\drillModule{edmOffline}{1.80}{-1.05}{3.60}{.85}{}
  {Offline preparation}{LLM descriptions}
\drillModule{edmHybrid}{1.80}{-2.30}{3.60}{1.05}{drillBlue}
  {Retrieval: Top-500}{BM25 + dense + HyDE\\+ citation expansion}
\drillModule{edmFirst}{1.80}{-3.55}{3.60}{.85}{drillBlue}
  {LambdaMART ranker}{EDM features, Top-100}
\drillModule{edmSecond}{1.80}{-4.80}{3.60}{1.05}{}
  {LLM-assisted reranking}{100 LLM pair scores\\Tree ranker $\to$ Top-10}
\drillModule{edmFinalLLM}{1.80}{-6.15}{3.60}{1.05}{drillBlue}
  {Final LLM evidence}{Top-10: 10 pair scores\\+ 1 collective judgment}
\drillModule{edmSelector}{1.80}{-7.40}{3.60}{.85}{drillBlue}
  {Pointwise tree selector}{Scores and threshold}
\node[drillBox,minimum width=3.60cm,minimum height=.45cm,fill=white]
  (edmOutput) at (1.80,-8.75) {Selected Article set};
\foreach \a/\b in {edmOffline/edmHybrid,edmHybrid/edmFirst,
  edmFirst/edmSecond,edmSecond/edmFinalLLM,
  edmFinalLLM/edmSelector,edmSelector/edmOutput}
  \draw[drillFlow] (\a.south)--(\b.north);

\drillModule{sharedPool}{8.25}{-1.15}{7.90}{1.05}{drillBlue}
  {Shared retrieval: Top-500}
  {BM25 + BGE-M3 + HyDE + citation expansion\\
   \textbf{Same candidate pool for both branches}}
\drillModule{controlRanker}{6.10}{-2.75}{3.60}{1.50}{drillBlue}
  {Control ranker}
  {EDM feature set\\Top-500 $\to$ Top-100}
\drillModule{proposedRanker}{10.40}{-2.75}{3.60}{1.50}{drillBlue}
  {Proposed ranker}
  {EDM feature set\\
   \drillBadge{drillLegalFill}{drillLegalInk}{Added legal evidence}\\
   Top-500 $\to$ Top-100}
\draw[drillFlow] (sharedPool.south -| controlRanker.north)--(controlRanker.north);
\draw[drillFlow] (sharedPool.south -| proposedRanker.north)--(proposedRanker.north);

\drillModule{alignment}{8.25}{-4.40}{7.90}{1.05}{fill=white}
  {Shared Top-10 for selection}
  {\textbf{Top-10 prefix of proposed Top-100}\\Same candidate IDs and ranking scores}
\draw[drillFlow] (controlRanker.south)--(controlRanker.south |- alignment.north);
\draw[drillFlow] (proposedRanker.south)--(proposedRanker.south |- alignment.north);

\drillModule{sharedFinal}{8.25}{-5.80}{7.90}{1.10}{drillBlue}
  {\textcolor{drillBlueInk}{Top-10 listwise LLM evidence}}
  {\drillBadge{drillAdaptFill}{drillAdaptInk}{One collective judgment}\\
   Qwen2.5-72B-Instruct}
\draw[drillFlow] (alignment.south)--(sharedFinal.north);

\drillModule{controlSelector}{6.10}{-7.40}{3.60}{1.40}{drillBlue}
  {Control selector}
  {EDM-based selection\\features}
\drillModule{proposedSelector}{10.40}{-7.40}{3.60}{1.40}{drillBlue}
  {Proposed selector}
  {EDM-based selection\\features\\
   \drillBadge{drillLegalFill}{drillLegalInk}{Added legal evidence}}
\draw[drillFlow] (sharedFinal.south -| controlSelector.north)--(controlSelector.north);
\draw[drillFlow] (sharedFinal.south -| proposedSelector.north)--(proposedSelector.north);
\node[drillBox,minimum width=3.60cm,minimum height=.45cm,fill=white]
  (controlOutput) at (6.10,-8.75) {Article set $S_q^{(0)}$};
\node[drillBox,minimum width=3.60cm,minimum height=.45cm,fill=white]
  (proposedOutput) at (10.40,-8.75) {Article set $S_q^{(1)}$};
\draw[drillFlow] (controlSelector.south)--(controlOutput.north);
\draw[drillFlow] (proposedSelector.south)--(proposedOutput.north);
\end{tikzpicture}%
\endgroup%
}
\caption{Published EDM~\cite{pham2025edm}, internal control and proposed method. \textcolor[HTML]{2D5E80}{Blue} marks EDM-derived stage designs, \textcolor[HTML]{8C5C17}{amber} simplified LLM evidence, and \textcolor[HTML]{246346}{green} added legal features. EDM also uses Qwen2.5-72B-Instruct. Arrows follow the comparison protocol. Both final selectors use the Top-10 prefix of the proposed Top-100 and shared LLM evidence.}
\label{fig:pipeline}
\end{figure}

Figure~\ref{fig:pipeline} distinguishes EDM inheritance from added legal evidence. Each branch ranks the shared Top-500 and keeps Top-100. Both final selectors use the first ten Articles and their scores from the proposed ranking, with shared listwise judgments and learner settings. Legal features and development thresholds differ. First-ranker configurations are not capacity-matched.

\subsection{Candidate retrieval and conventional evidence}
We combine BM25~\cite{robertson2009bm25} with frozen BGE-M3 question scores~\cite{chen2024bgem3}. A HyDE passage supplies a second dense query view. Local citations provide bounded expansion (at most 80 additions per view). Merged candidates form $C_q$, with at most 500 unique Article IDs. Neither the legal feature vector nor the final selector adds candidates beyond its input pool.

Both first-ranker branches use the 22 conventional feature definitions in EDM~\cite{pham2025edm}. The vector $x_q(a)$ includes a frozen BGE reranker score, dense and HyDE scores, auxiliary Vietnamese embedding similarities, reciprocal rank fusion~\cite{cormack2009rrf}, a normalized pool score, corpus positions, length and form cues, and compatibility fields. For added legal evidence $\phi_q(a)\in\mathbb R^{27}$, the paired representations are
\begin{equation}
 h_q^{\rm ctrl}(a)=x_q(a)\in\mathbb R^{22},\qquad
 h_q^{\rm legal}(a)=x_q(a)\oplus\phi_q(a)\in\mathbb R^{49}.
 \label{eq:features}
\end{equation}

\Needspace{8\baselineskip}
\subsection{Features informed by legal domain knowledge}
Table~\ref{tab:features} explains three feature groups: identity and local form (1--10, blue), same-law support (11--13, green), and citation/lineage support (14--27, amber). They address ambiguity left by similarity. Incoming features summarize Articles pointing to the candidate, while outgoing features summarize those it cites. Directed citation relations motivate this distinction~\cite{colombo2025legissearch}.

\begingroup
\definecolor{legalidentity}{HTML}{E7EFF8}
\definecolor{legalsamelaw}{HTML}{E5F1E8}
\definecolor{legalcitation}{HTML}{FFF0D9}
\makeatletter
\def\LT@hline{\noalign{\penalty\@M\hrule height\arrayrulewidth\penalty\z@}}
\makeatother
\footnotesize
\setlength{\tabcolsep}{4pt}
\setlength{\LTleft}{0pt}
\setlength{\LTright}{0pt}
\setlength{\LTcapwidth}{\linewidth}
\renewcommand{\arraystretch}{1.07}
\begin{longtable}{>{\raggedleft\arraybackslash}p{0.45cm}>{\columncolor{tableband}\raggedright\arraybackslash}p{3.15cm}>{\raggedright\arraybackslash}p{\dimexpr\linewidth-0.45cm-3.15cm-6\tabcolsep\relax}}
\caption{Features informed by legal domain knowledge.}\label{tab:features}\\
\hline
\textbf{ID} & \textbf{Feature} & \textbf{Explanation}\\
\hline
\endfirsthead
\caption[]{Features informed by legal domain knowledge (continued).}\\
\hline
\textbf{ID} & \textbf{Feature} & \textbf{Explanation}\\
\hline
\endhead
\multicolumn{3}{r}{\footnotesize Continued on the next page}\\
\endfoot
\hline
\endlastfoot
1 & \cellcolor{legalidentity}Article/alias match & Query numbering distinguishes an intended provision or its introducing Article from topical neighbors.\\\hline
2 & \cellcolor{legalidentity}Compatible Clause cue & A Clause cue marks a request for a specific condition within a compatible Article.\\\hline
3 & \cellcolor{legalidentity}Compatible Point cue & A Point cue marks finer targeting of an enumerated case or exception within a compatible Article.\\\hline
4 & \cellcolor{legalidentity}Instrument-ID mention & An instrument ID disambiguates statutes that reuse Article numbers or discuss similar topics.\\\hline
5 & \cellcolor{legalidentity}Article coreference & Article self-reference signals that an isolated text match may omit necessary surrounding context.\\\hline
6 & \cellcolor{legalidentity}Clause coreference & Clause self-reference flags conditions requiring context from another part of the candidate.\\\hline
7 & \cellcolor{legalidentity}Point coreference & Point self-reference highlights qualifications that depend on finer enumerated context.\\\hline
8 & \cellcolor{legalidentity}Log Clause count & Clause count approximates procedural or conditional breadth. Log encoding represents it compactly.\\\hline
9 & \cellcolor{legalidentity}Log Point count & Point count captures finer cases and exceptions beyond Clause breadth, using the same log encoding.\\\hline
10 & \cellcolor{legalidentity}Amendment-action match & Matching change-action types distinguishes a requested modification from ordinary discussion of the rule.\\\hline
11 & \cellcolor{legalsamelaw}Same-law fraction & The fraction of pool candidates from the same instrument measures support without individual scores.\\\hline
12 & \cellcolor{legalsamelaw}Same-law maximum & One strongly scored same-law Article may identify a useful instrument for a weaker companion.\\\hline
13 & \cellcolor{legalsamelaw}Same-law mean & Mean same-law strength separates broad support from one unusually strong match.\\\hline
14 & \cellcolor{legalcitation}Local incoming maximum & The strongest retrieved citer may reveal a candidate needed to interpret another provision.\\\hline
15 & \cellcolor{legalcitation}Local incoming sum & Summed citer scores represent cumulative support from several references, complementing a single strongest citation.\\\hline
16 & \cellcolor{legalcitation}Local incoming mean & Mean citer strength separates typical support from citation volume and exceptional neighbors.\\\hline
17 & \cellcolor{legalcitation}Local outgoing maximum & A strong cited provision may connect qualifying text to the underlying rule relevant to the query.\\\hline
18 & \cellcolor{legalcitation}Local outgoing mean & Average cited-neighbor strength reduces reliance on one unusually strong outgoing connection.\\\hline
19 & \cellcolor{legalcitation}Letter-suffix introduction & Letter-suffix introductions identify inserted legal units that ordinary corpus-position numbering may miss.\\\hline
20 & \cellcolor{legalcitation}Introduced-ID query match & A query naming an introduced unit may require the Article that contains its introduction.\\\hline
21 & \cellcolor{legalcitation}Typed outgoing maximum & Extended outgoing links capture cross-instrument context beyond local references.\\\hline
22 & \cellcolor{legalcitation}Typed incoming maximum & Extended incoming links capture strong support from candidates in other instruments.\\\hline
23 & \cellcolor{legalcitation}Amendment outgoing maximum & An amendment-tagged candidate may modify the underlying rule indicated by its strongest target.\\\hline
24 & \cellcolor{legalcitation}Amendment incoming maximum & A strong amendment-tagged citer may indicate that the candidate rule has been changed.\\\hline
25 & \cellcolor{legalcitation}Guidance outgoing maximum & A guidance-tagged candidate may detail or implement the rule indicated by its strongest target.\\\hline
26 & \cellcolor{legalcitation}Guidance incoming maximum & A strong guidance-tagged citer may supply implementation context for the candidate rule.\\\hline
27 & \cellcolor{legalcitation}Introduced-by maximum & An incoming introducer naming the candidate's indexed unit provides more specific evidence than amendment wording.\\
\end{longtable}
\endgroup

\noindent Article numbers use corpus positions or detected introduced aliases. Clause and Point cues check parent-Article compatibility without resolving subunits. Coreference flags mark unresolved self/above references. Line-start Clause/Point counts use $\log(1+n)$ to compress their numerical range while preserving order and mapping zero to zero. This is an encoding choice, not a requirement of tree learners. We recompute the legal feature vector $\phi_q(a)$ using BGE similarity on Top-500 and first-ranker scores on the shared Top-10. Only retrieved neighbors contribute, with pool min-max normalization and zero for missing summaries. Same-law summaries include the candidate. Local outgoing edges exclude sources matching the amendment-action pattern. Typed links connect every Article of an explicitly named different instrument. Action tags label citing Articles, not verified edges. Feature~19 requires a letter suffix, while Feature~27 matches an incoming introduced alias to the candidate's indexed number. These cues do not verify legislative histories or effective dates.

\subsection{From ten candidates to an answer set}
The shared list $T_q$ is the first ten entries of the proposed Top-100, with no intervening reranker. One Qwen2.5-72B-Instruct~\cite{qwen2025qwen25} listwise judgment reads the question, IDs, and up to 3,000 characters per Article, returning a score for every candidate and a proposed minimal index set. The ten-dimensional conventional vector $z_q(a)$ adapts EDM's score fusion: cross-encoder, first-ranker, LLM, and fusion scores, their four normalized counterparts, the LLM-selected flag, and reciprocal first-ranker position. The proposed selector adds $\phi_q(a)$ to $z_q(a)$.

Both selectors use LightGBM~\cite{ke2017lightgbm} with objective \texttt{rank\_xendcg}~\cite{bruch2021xendcg}. Their raw scores are $u_q^{(b)}(a)=f_b(h)$, where $h=z_q(a)$ or $z_q(a)\oplus\phi_q(a)$. Min-max normalization within $T_q$ produces relative scores, not calibrated probabilities. Selection uses thresholds $\tau_0=0.29$ and $\tau_1=0.30$, with a top-one fallback if none passes:
\begin{equation}
 \begin{gathered}
 s_q^{(b)}(a)=\frac{u_q^{(b)}(a)-\min_{c\in T_q}u_q^{(b)}(c)}
 {\max\{10^{-8},\max_{c\in T_q}u_q^{(b)}(c)-\min_{c\in T_q}u_q^{(b)}(c)\}},\\[3pt]
 S_q^{(b)}=\operatorname{Top}_{5}\{a\in T_q:s_q^{(b)}(a)\geq\tau_b\}.
 \end{gathered}
 \label{eq:select}
\end{equation}

\subsection{Vietnamese prompts}
The panels give the shared system message and complete user prompts, with dynamic input blocks defined below.

\begingroup
\small
\setlength{\fboxsep}{4pt}
\setlength{\fboxrule}{0.35pt}
\setlength{\parindent}{0pt}
\setlength{\parskip}{0pt}
\noindent\textbf{Shared system message.} Chỉ thực hiện legal article retrieval và chỉ xuất JSON hợp lệ.

\smallskip
\noindent\fcolorbox{black!35}{softblue}{%
\begin{minipage}{\dimexpr\linewidth-2\fboxsep-2\fboxrule\relax}
\small\raggedright
\textbf{HyDE user message}\par\smallskip
\noindent Viết một đoạn quy phạm pháp luật giả định 60--120 từ cho mỗi câu hỏi để mở rộng truy vấn. Nêu chủ thể, hành vi, điều kiện và từ ngữ pháp lý có thể xuất hiện; không trả lời, không kết luận và không bịa số Điều/số hiệu.\par
\noindent Chỉ xuất JSON: \texttt{\{"items":[\{"id":"...","text":"..."\}]\}}. Giữ đủ ID và thứ tự.\par
\smallskip
\noindent\texttt{\{rendered\}}\par
\end{minipage}}
\par

\medskip
\noindent\fcolorbox{black!35}{softgreen}{%
\begin{minipage}{\dimexpr\linewidth-2\fboxsep-2\fboxrule\relax}
\small\raggedright
\textbf{Listwise user message}\par\smallskip
\noindent Bạn là bộ chọn Điều luật cho tác vụ truy hồi văn bản pháp luật Việt Nam.\par
\smallskip
\noindent Nhiệm vụ duy nhất: đánh giá Điều nào trong danh sách ứng viên chứa căn cứ pháp lý trực tiếp và cần thiết cho câu hỏi. Không trả lời câu hỏi, không suy luận đáp án cuối cùng và không viết giải thích. Hãy phân tích nội bộ lĩnh vực pháp luật, chủ thể, hành vi, điều kiện/ngoại lệ và loại quy phạm mà câu hỏi cần trước khi chấm. Đặc biệt phân biệt các Điều có câu chữ giống nhau nhưng thuộc lĩnh vực hoặc thủ tục khác. Không chọn Điều chỉ vì cùng luật, có từ khóa giống nhau, hoặc chỉ dẫn sang Điều khác. Nếu nhiều Điều thực sự bổ sung điều kiện/ngoại lệ cho nhau thì chọn đủ, nhưng tập chọn phải tối thiểu.\par
\smallskip
\noindent Chấm MỖI ứng viên theo rubric:\par
\noindent - 0.90--1.00: chứa trực tiếp quy phạm quyết định câu hỏi;\par
\noindent - 0.70--0.89: căn cứ bổ sung cần thiết, không thể bỏ;\par
\noindent - 0.30--0.69: liên quan nhưng không đủ/không cần thiết;\par
\noindent - 0.00--0.29: chỉ giống từ khóa hoặc không liên quan.\par
\noindent Sau đó chọn tập chỉ số tối thiểu nhưng đầy đủ. Chỉ xuất đúng một JSON object theo schema:\par
\noindent\texttt{\{"scores":[\{"index":0,"score":0.0\}],"selected":[0]\}}\par
\smallskip
\noindent CÂU HỎI:\par
\noindent\texttt{\{query\}}\par
\smallskip
\noindent ỨNG VIÊN:\par
\noindent\texttt{\{blocks\}}\par
\end{minipage}}
\par\smallskip
\noindent Input blocks are separated by blank lines. HyDE batches up to eight questions. Listwise selection uses ten candidates with \texttt{budget=3000} characters:
\par\smallskip
\noindent\begin{minipage}[t]{.48\linewidth}
\textbf{\texttt{\{rendered\}}: question block}\par\smallskip
\ttfamily
[ID \{query.key\}]\\
\{query.text.replace('\_', ' ')\}
\end{minipage}\hfill
\begin{minipage}[t]{.48\linewidth}
\textbf{\texttt{\{blocks\}}: candidate block}\par\smallskip
\ttfamily
[ỨNG VIÊN \{index\}]\\
ID văn bản: \{article.law\_id\}\\
ID Điều: \{article.article\_id\}\\
Nội dung:\\
\{article.text[:budget]\}
\end{minipage}
\par
\endgroup

\clearpage

\begingroup
\setlength{\intextsep}{6pt}
\setlength{\textfloatsep}{6pt}
\section{Experimental Setup}
\subsection{Models and selection policy}
HyDE uses Qwen2.5-72B-Instruct at temperature 0.20, batching up to eight questions. Encoders and reranker remain frozen. Table~\ref{tab:settings} reports the tree hyperparameters.

\begin{table}[H]\centering\small
\caption{LightGBM hyperparameters. Child samples are question/Article pairs. First-ranker configurations are not capacity-matched.}
\label{tab:settings}\setlength{\tabcolsep}{4.5pt}
\begin{tabular}{l>{\columncolor{tableband}}r r>{\columncolor{tableband}}r}\toprule
\textbf{Setting} & \textbf{Control ranker} & \textbf{Proposed ranker} & \textbf{Both selectors}\\\midrule
Trees & 700 & 600 & 400\\\hline
Max. leaves / depth & 7 / 4 & 15 / 6 & 31 / 5\\\hline
Minimum child samples & 20 & 40 & 40\\\hline
Learning rate & 0.07325 & 0.06467 & 0.03000\\\hline
$L_2$ regularization & 2.75643 & 0.31089 & 1.00000\\\bottomrule
\end{tabular}
\end{table}

All feature configurations, learner hyperparameters, candidate budgets, and thresholds are chosen on development data. Five-fold out-of-fold ranker predictions create score features for training the final LightGBM selector: each training question is scored by a model fitted on the other four folds. The folds are used only to construct training inputs. Rankers are then refitted on all training questions for development/test prediction. Seed-0 development choices remain fixed across seeds 1--4.

\subsection{Seeds, shared judgments, and reporting}
HyDE and trees use seeds 0--4. Listwise Qwen currently uses seed 0 and temperature 0 due to computational resource constraints. Both selectors share candidate-aligned judgments within each run. Top-10 lists may vary through seeded upstream components, so SD reflects run variability under fixed Qwen decoding.

Where shown, $\pm$ denotes sample SD over five seeds. Query-level wins, ties, and losses compare five-seed mean $F_{2,q}$. Performance differences use percentage points (pp), with differences and totals computed before rounding.

Two-sided paired randomization uses $10^6$ permutations, with each question swapped consistently across the five seeds. The statistic is the five-seed mean difference, with Eq.~\eqref{eq:f2} recomputed per seed. Benjamini--Hochberg correction~\cite{urbano2019significance} jointly adjusts R@10, R@100, precision, recall, and official $F_2$.

\section{Experimental Results}
\subsection{Ordering within a fixed candidate pool}
In the fixed Top-500 pool, R@10 increases by 0.95 points and R@100 by 0.77 (Table~\ref{tab:rank}, unrounded differences). Only R@100 is significant after BH correction. Ranker hyperparameters differ, so these gains cannot be attributed solely to legal features.

\begin{table}[H]\centering\small
\caption{Test recall (\%, five-seed mean $\pm$ sample SD). $\dagger$: BH-adjusted $p<0.05$ versus control.}
\label{tab:rank}\setlength{\tabcolsep}{10pt}
\begin{tabular}{l>{\columncolor{tableband}}c c}\toprule
\textbf{Configuration} & \textbf{R@10} & \textbf{R@100}\\\midrule
Control: 22 conventional & $87.33\pm0.44$ & $96.75\pm0.21$\\\hline
Proposed: 22 + 27 legal & $\best{88.27\pm0.74}$ & $\best{97.52\pm0.15}$\sigfive\\\bottomrule
\end{tabular}
\end{table}

\subsection{Selecting the relevant Article set}
With shared learner settings, candidates, and listwise evidence, but branch-specific development thresholds, precision gains 1.99 points, recall 0.75, and official $F_2$ 1.14 (Table~\ref{tab:final}). All three differences pass BH correction ($p_{\rm BH}=0.00078$ for $F_2$). Mean per-query $F_2$ gains 1.33 because its averaging differs from the official metric. All five runs improve final $F_2$. Output size falls in four and is unchanged in one (Table~\ref{tab:seeds}). The gain does not follow from returning larger sets.

\begin{table}[H]\centering\small
\caption{Per-seed differences (proposed minus control) in recall and official $F_2$ (pp). Set sizes are mean selected Articles per test question.}
\label{tab:seeds}\setlength{\tabcolsep}{7pt}
\begin{tabular}{c>{\columncolor{tableband}}c c>{\columncolor{tableband}}c c}\toprule
\textbf{Seed} & \textbf{$\Delta$R@10} & \textbf{$\Delta F_2$} & \textbf{Mean $|S|$ control} & \textbf{Mean $|S|$ proposed}\\\midrule
0 & +0.94 & +0.76 & 2.08 & 2.04\\\hline
1 & +1.56 & +2.12 & 2.05 & 2.05\\\hline
2 & +0.87 & +0.77 & 2.02 & 2.00\\\hline
3 & +0.85 & +1.51 & 2.07 & 2.04\\\hline
4 & +0.52 & +0.55 & 2.02 & 1.96\\\bottomrule
\end{tabular}
\end{table}

\subsection{VLSP results in context}
Table~\ref{tab:vlsp} compares our method with the three leading systems in the DRiLL competition. Our proposed five-run mean is 0.30 percentage points above EDM's reported $F_2$, with lower precision and higher recall. This is a descriptive cross-system comparison. Configurations and reporting protocols differ, and paired predictions for other teams are unavailable.

\begin{table}[H]\centering\small
\caption{Private-test precision, recall and official $F_2$ (\%, five-seed mean $\pm$ sample SD for our systems). Top-three scores follow~\cite{vuong2025drill}. For proposed versus control, $\dagger/\ddagger$ mark BH-adjusted $p<0.05/0.01$.}
\label{tab:vlsp}\label{tab:final}\setlength{\tabcolsep}{6pt}
\begin{tabular}{l>{\columncolor{tableband}}l l>{\columncolor{tableband}}l}\toprule
\textbf{System} & \textbf{Precision} & \textbf{Recall} & \textbf{$F_2$}\\\midrule
ducanger~\cite{hieu2025twostage} & $50.97$ & $76.53$ & $69.55$\\\hline
unknow\_123~\cite{hoai2025chunking} & $62.22$ & $71.81$ & $69.66$\\\hline
EDM (edmmm)~\cite{pham2025edm} & $67.73$ & $73.94$ & $72.61$\\\hline
Control (this study) & $57.13\pm0.80$ & $76.68\pm0.51$ & $71.77\pm0.47$\\\hline
Proposed (this study) & $59.11\pm0.35$\sigone & $77.43\pm1.03$\sigfive & $72.91\pm0.68$\sigone\\\bottomrule
\end{tabular}
\end{table}

\endgroup

\section{Analysis}
\subsection{Feature importance}
\begin{figure}[!t]\centering
\includegraphics[width=\linewidth]{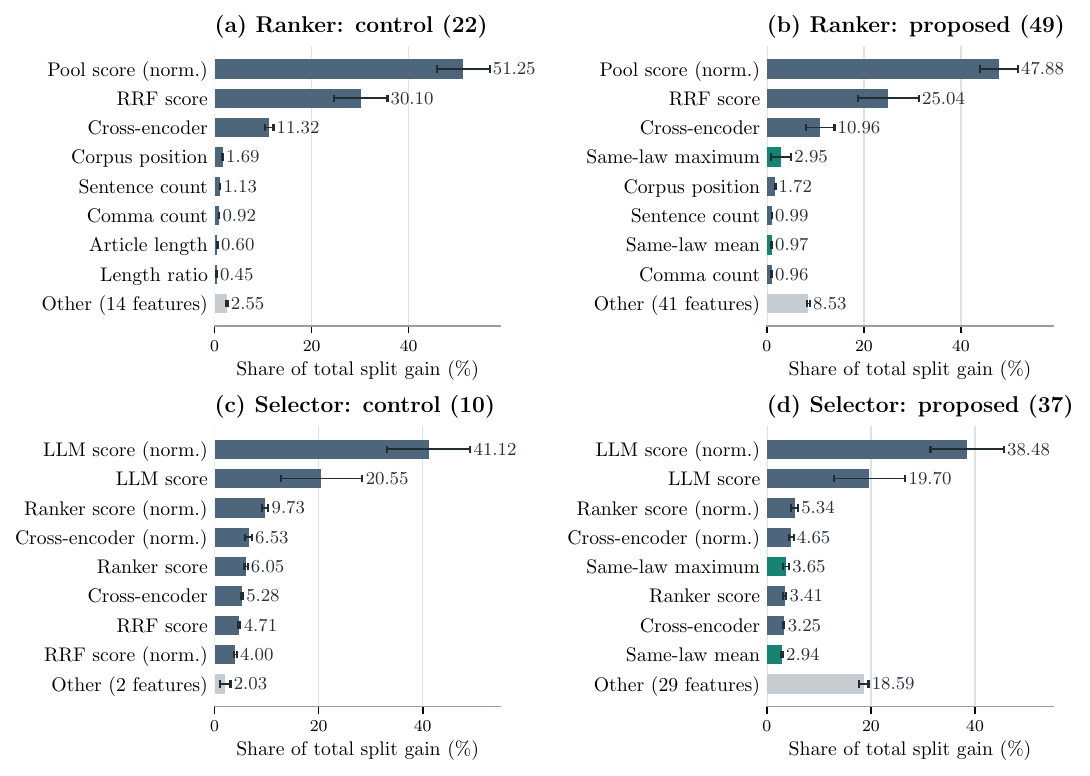}
\caption{Training split-gain importance for all four model families: eight leading features and the remainder. Bars show mean within-model split-gain shares and whiskers sample SD across five seeds. Blue denotes conventional features, green same-law features, and gray other features.}
\label{fig:importance}
\end{figure}

\Needspace{6\baselineskip}
Figure~\ref{fig:importance} reports split-gain importance in the fitted models, following LightGBM's definition~\cite{lightgbm2025gain}. For feature $j$ in model $m$, normalized importance is
\begin{equation}
 I_{m,j}=100\,\frac{\sum_{t:\,\operatorname{feature}(t)=j}\operatorname{gain}(t)}
 {\sum_t\operatorname{gain}(t)}.
 \label{eq:importance}
\end{equation}
Here $t$ indexes tree splits and $\operatorname{gain}(t)$ is their estimated training-objective reduction. Group sums are normalized by all-feature gain within each model before five-seed summarization. These shares describe model use, not test $F_2$ changes or ablation effects. Correlated inputs may substitute for one another, raw totals are incomparable across models/objectives, and groups contain 10, 3, and 14 features.

Normalized pool scores and fusion dominate ranker gain. Normalized and raw LLM scores dominate selector gain. Added legal features contribute 7.53\% and 18.20\%, respectively, with same-law support the largest added group (Table~\ref{tab:gain}).

\begingroup
\setlength{\intextsep}{6pt}
\begin{table}[H]\centering\small
\caption{Feature importance in proposed models: share of total training split gain across all features (\%, mean $\pm$ sample SD over five seeds).}
\label{tab:gain}\setlength{\tabcolsep}{7pt}
\begin{tabular}{l>{\columncolor{tableband}}c c}\toprule
\textbf{Added feature group} & \textbf{Ranker (\%)} & \textbf{Selector (\%)}\\\midrule
Identity / local form (1--10) & $0.50\pm0.01$ & $2.87\pm0.16$\\\hline
Same-law support (11--13) & $4.65\pm2.11$ & $8.04\pm0.57$\\\hline
Citation / lineage (14--27) & $2.38\pm0.19$ & $7.28\pm0.32$\\\bottomrule
\end{tabular}
\end{table}
Of the selector's 7.28\% citation/lineage share (Table~\ref{tab:gain}), 4.83\% comes from local references (14--18) and 1.32\% from introduction/amendment/guidance (19--20, 23--27). The group is therefore not exclusively evidence for lineage. Same-law maximum leads the added features (2.95\% ranker, 3.65\% selector). Features 1--4 and 20 have zero gain in every proposed model. Zero gain does not establish irrelevance under other encodings.

\FloatBarrier

\subsection{Aggregate analysis and concrete cases}
\begin{table}[H]\centering\small
\caption{Mean per-query $F_{2,q}$ (\%), averaged over seeds and questions within each gold-size group. $n$ counts test questions. $\Delta\overline{F_{2,q}}$ is proposed minus control (pp). W/T/L count wins/ties/losses using each question's five-seed mean.}
\label{tab:cardinality}\setlength{\tabcolsep}{5pt}
\begin{tabular}{c>{\columncolor{tableband}}r r>{\columncolor{tableband}}r r>{\columncolor{tableband}}r r>{\columncolor{tableband}}r}\toprule
\textbf{$|G_q|$} & \textbf{$n$} & \textbf{Control} & \textbf{Proposed} & \textbf{$\Delta\overline{F_{2,q}}$} & \textbf{W} & \textbf{T} & \textbf{L}\\\midrule
1 & 474 & 71.49 & 71.91 & +0.42 & 166 & 217 & 91\\\hline
2 & 121 & 61.34 & 65.79 & +4.45 & 61 & 42 & 18\\\hline
$\geq3$ & 32 & 43.51 & 46.40 & +2.89 & 12 & 11 & 9\\\bottomrule
\end{tabular}
\end{table}
\endgroup

Table~\ref{tab:cardinality} groups outcomes by distinct gold-set size. The largest gain is 4.45 points for the 121 two-Article questions, versus 0.42 for single-Article questions. Larger sets remain difficult, with proposed per-query $F_2$ of 46.40\%.

For the 121 two-Article questions across five seeds (605 paired selections), we count Article occurrences added or removed relative to control. The proposed branch adds 74 gold occurrences and removes 11, while removing 115 non-gold occurrences and adding 95. Complete gold coverage rises from 222 to 283 paired selections. However, among non-gold Articles from a gold instrument, 46 are removed and 67 added: the aggregate does not support a claim of improved same-law distractor pruning. Counts reuse the same questions across seeds and are not independent trials or a causal decomposition of $F_2$.

The DRiLL cases~\cite{vuong2025drill} give original questions, English translations, legal evidence, and five-seed selections. $G$, $C$, and $P$ denote gold, control, and proposed corpus \texttt{aid} sets. Bold phrases highlight benchmark wording. The cases do not isolate individual feature effects.

\FloatBarrier
\newpage
\begingroup
\setlength{\intextsep}{6pt}
\begin{table}[H]
\noindent\normalsize Table~\ref{tab:case8360} presents a dossier question whose gold set combines a base provision with an amendment. The proposed method improves on control by recovering the missing gold amendment in four of five runs. The amendment changes both the required number of dossier sets and the submission format, making it complementary to the base provision.
\par\vspace{5pt}
\centering\small
\emergencystretch=1.5em
\setlength{\tabcolsep}{4pt}
\renewcommand{\arraystretch}{1.08}
\caption{Improvement over control: companion recovery (question 8360).}
\label{tab:case8360}
\begin{tabularx}{\linewidth}{>{\columncolor{tableband}\raggedright\arraybackslash}p{1.8cm}>{\raggedright\arraybackslash}X}
\hline
\textbf{Field} & \textbf{Question, evidence and outcome}\\\hline
Vietnamese & \emph{Hồ sơ đăng ký bình chọn sản phẩm công nghiệp nông thôn tiêu biểu cấp huyện bao gồm những gì?}\\\hline
English & What does the application dossier for district-level selection of outstanding rural industrial products include?\\\hline
Legal text & \textbf{Base:} Circular~26/2014/TT-BCT, Art.~12 (\texttt{49460}), lists application documents and requires ``chuẩn bị \textbf{01 bộ hồ sơ}'' (prepare one dossier).
\par\smallskip
\textbf{Amendment:} Circular~14/2018/TT-BCT, Art.~1(6) (\texttt{49474}), amends Art.~12(1): ``gửi \textbf{02 (hai) bộ hồ sơ} (kèm theo bản \textbf{hồ sơ điện tử})''. It requires two dossier sets plus an electronic dossier.\\\hline
Selections & $G=\{49460,49474\}$. Control returns $C=\{49460\}$ in all five runs. Proposed returns $P=G$ in seeds~0, 1, 2 and~4, and $P=\{49460\}$ in seed~3.\\
\hline
\end{tabularx}
\end{table}
\begin{table}[H]
\noindent\normalsize Table~\ref{tab:case411} presents an inheritance question whose gold Articles provide complementary rules on the scope of statutory succession and equal shares. Compared with control, the proposed method removes the same-law non-gold Article in all five runs. Art.~650 remains missing in every run, so the improvement leaves the benchmark evidence incomplete.
\par\vspace{5pt}
\centering\small
\emergencystretch=1.5em
\setlength{\tabcolsep}{4pt}
\renewcommand{\arraystretch}{1.08}
\caption{Partial improvement over control: distractor removal (question 411).}
\label{tab:case411}
\begin{tabularx}{\linewidth}{>{\columncolor{tableband}\raggedright\arraybackslash}p{1.8cm}>{\raggedright\arraybackslash}X}
\hline
\textbf{Field} & \textbf{Question, evidence and outcome}\\\hline
Vietnamese & \emph{Con trong giá thú và con ngoài giá thú có được hưởng di sản thừa kế như nhau hay không?}\\\hline
English & Are children born within and outside marriage entitled to equal inheritance?\\\hline
Legal text & All three Articles belong to Civil Code~91/2015/QH13.
\par\smallskip
\textbf{Scope:} Art.~650 (\texttt{53420}) specifies when statutory succession applies. Art.~650(1)(a), ``\textbf{Không có di chúc}'' (there is no will), is one such circumstance.
\par\smallskip
\textbf{Equal shares:} Art.~651 (\texttt{53421}) identifies heirs. Art.~651(2) states ``Những người thừa kế \textbf{cùng hàng} được hưởng phần di sản \textbf{bằng nhau}.'' Heirs in the same rank receive equal shares.
\par\smallskip
\textbf{Distractor:} Art.~652 (\texttt{53422}) concerns inheritance by representation for grandchildren and great-grandchildren.\\\hline
Selections & $G=\{53420,53421\}$. In all five runs, $C=\{53421,53422\}$ and $P=\{53421\}$.\\
\hline
\end{tabularx}
\end{table}
\endgroup
\newpage

\begin{table}[H]
\noindent\normalsize Table~\ref{tab:case7758} presents a record-retention question whose gold set contains a base provision and its replacement. Control returns both gold Articles in all five runs, whereas the proposed method drops the amendment in two of five runs. The replacement rule specifies electronic storage while still requiring selected records to be retained in writing.
\par\vspace{5pt}
\centering\small
\emergencystretch=1.5em
\setlength{\tabcolsep}{4pt}
\renewcommand{\arraystretch}{1.08}
\caption{Failure relative to control: amendment loss (question 7758).}
\label{tab:case7758}
\begin{tabularx}{\linewidth}{>{\columncolor{tableband}\raggedright\arraybackslash}p{1.8cm}>{\raggedright\arraybackslash}X}
\hline
\textbf{Field} & \textbf{Question, evidence and outcome}\\\hline
Vietnamese & \emph{Những tài liệu đánh giá xếp loại chất lượng công chức nào sẽ được lưu trữ?}\\\hline
English & Which documents on civil servants' performance assessment and classification will be archived?\\\hline
Legal text & \textbf{Base:} Decree~90/2020/NĐ-CP, Art.~22 (\texttt{28455}), lists assessment records retained in personnel files.
\par\smallskip
\textbf{Amendment:} Decree~48/2023/NĐ-CP, Art.~1(2) (\texttt{58296}), replaces Art.~22. Records ``được lưu giữ bằng hình thức \textbf{điện tử}'' (are retained electronically). Items~2 and~4 (``khoản 2 và khoản 4'') of the replacement Art.~22 also ``lưu bằng \textbf{văn bản}'' (remain in writing). These items concern assessment forms and conclusions/notices.\\\hline
Selections & $G=\{28455,58296\}$. Control returns $C=G$ in all five runs. Proposed returns $P=G$ in seeds~0, 1 and~2, and $P=\{28455\}$ in seeds~3 and~4.\\
\hline
\end{tabularx}
\end{table}

\FloatBarrier

\subsection{Transfer and inference work}
Reusing instrument aggregation and directional summaries requires mapped units, stable IDs, adapted Article/Clause/Point patterns and action lexicons, and rebuilt reference edges. Numbering differences, consolidated texts, sparse citations, and broad references can reduce coverage or add noise. Transfer is motivated by statutory-graph retrieval~\cite{louis2023finding} but remains unvalidated.

EDM additionally uses offline law/Article descriptions, Top-100 LLM pair scores, a second learned ranker, and ten pairwise plus one collective final judgment~\cite{pham2025edm}. Our path uses HyDE and one collective Top-10 judgment. The nominal online work is therefore up to $1+100+10+1=112$ versus $1+1=2$ logical LLM evaluations. Final selection alone is 11 versus 1. Batching, caching, retries, token lengths, and neural reranking are outside these counts, so they do not establish API-request, cost, or latency ratios.

\section{Conclusion}
DRiLL cases demonstrate structural relationships and legislative lineage. Split-gain importance and paired selection changes illustrate their use by our retrieval method. Shared-input final selection improves official $F_2$ from 71.77\% to 72.91\%, especially for two-Article questions. These findings motivate cross-dataset validation and alternative retrieval, representation-learning, or multi-Article reasoning methods. They establish neither the individual benefit of every feature nor the necessity of LightGBM, hand-crafted scores, or explicit graphs.

\Needspace{24\baselineskip}
\section*{Limitations and Ethical Considerations}
\noindent\textbf{Scope.} Evaluation covers DRiLL only. Cross-dataset validation, expert review of parsed links, and matched latency/cost measurements remain future work.

\noindent\textbf{Attribution.} We have not yet performed ablations of individual legal features or their subgroups. First-ranker capacities differ without a matched-capacity comparison. Final selectors share the proposed Top-10 but use different development thresholds. Gain describes fitted-tree use, not causal importance.

\noindent\textbf{Legal use.} Heuristic links and amendment cues do not establish current legal validity or complete version histories. Retrieved Articles require qualified review of operative provisions and effective dates. Hypothetical passages are not legal advice. Confidential queries require safeguards when using external LLM services.

\section*{Acknowledgements}
This research was supported by The VNUHCM-University of Information Technology's Scientific Research Support Fund.

\begingroup
\interlinepenalty=10000
\bibliographystyle{splncs04}
\bibliography{references}
\endgroup
\end{document}